\documentclass[aps,pra,twocolumn,amsmath,amssymb,10pt]{revtex4-2}
\usepackage{amsmath,amssymb,graphicx,color,hyperref}
\usepackage{bm}
\usepackage{amsfonts}

\begin{document}

\title{Compatibility of transport effects in non-Hermitian nonreciprocal systems}

\author{Hamed Ghaemi-Dizicheh}
\affiliation{Department of Physics, Lancaster University, Lancaster, LA1 4YB, United Kingdom}
\author{Henning Schomerus}
\affiliation{Department of Physics, Lancaster University, Lancaster, LA1 4YB, United Kingdom}

\begin{abstract}
Based on a general transport theory for non-reciprocal non-Hermitian systems and a topological model that encompasses a wide range of previously studied examples, we (i) provide conditions for effects such as reflectionless and transparent transport, lasing, and coherent perfect absorption, (ii) identify which effects are compatible and linked with each other, and (iii) determine by which levers they can be tuned independently.
For instance, the directed amplification inherent in the non-Hermitian skin effect does not enter the spectral conditions for reflectionless transport, lasing, or coherent perfect absorption, but allows to adjust the transparency of the system. In addition, in the topological model the conditions for reflectionless transport
depend on the topological phase, but those for coherent perfect absorption do not.
This then allows us to establish a number of distinct transport signatures of non-Hermitian, nonreciprocal, and topological behaviour,
in particular (I) reflectionless transport in a direction that depends on the topological phase, (II)
invisibility coinciding with the skin-effect phase transition of topological edge states,
and (III)
coherent perfect absorption in a system that is transparent when probed from one side.

\end{abstract}
\maketitle

\section{Introduction}

Effectively non-Hermitian models have a long tradition in the description of states with a finite life time, with applications ranging from scattering resonances  over quasiparticle dephasing to classical wave propagation with gain and loss  \cite{Moi11,Cao15,ElG18}.
Over the last few years, these endeavours have received substantial  impetus  by the realization that non-Hermitian physics can equip existing topological states with unique physical features,
and also function as a source of topological effects in themselves \cite{Rud09,Esa11,Sch13,Zhu14,Pol15,Mal15,Lee16,Ley17,Ni18,Gon18,Car18,Zho18,Mal18,Kaw19,Mos20a,Ber21}. A particularly prominent manifestation is the non-Hermitian skin effect, in which the bulk states become localized at an edge of a finite system, resulting in a behaviour that is drastically different from its periodic counterpart \cite{Yao18,Xio18,Kun18,Mar18,Yao18b,Lon19,Yok19,Son19,Lee19,Jin19,Oku20,Yi20,Yuc20,Bor20}. While traditional non-Hermitian physics is mostly captured in imaginary scalar potentials that describe local gain and loss,
the non-Hermitian skin effect relies on imaginary vector potentials \cite{Hat96}, making  the system nonreciprocal. Recent experiments on electronic systems and mechanical robotic metamaterials have shown how this can be achieved in practice by inducing asymmetrical couplings between discrete components of an active system  \cite{Bra19,Gha20,Hof20,Hel20}. The unidirectional distortion can induce a dynamical phase transition, which can be utilized for unidirectional amplification and sensing applications  \cite{McD18,Sch20,Wan20,McD20,Bud20}.

This rich diversity of phenomena leaves the natural question of how the different effects are interlinked---specifically, whether different effects are compatible with each other, and can be achieved simultaneously by tuning suitable parameters, possibly facilitated by symmetry or topology.
In reciprocal systems, this line of thought has already proved  highly fruitful, as is testified  by the example of a laser-absorber---a lasing device that simultaneously can absorb a prescribed coherent signal at the lasing frequency, which is facilitated by parity-time symmetry \cite{Lon10,Cho11}.

Here, we address these connections for the general, nonreciprocal and non-Hermitian case. Aiming at a description that is physical and flexible, we adopt the unifying perspective of transport, which has been instrumental to identify the specific  signatures of individual physical effects in Hermitian \cite{Bee97,Bee15} and non-Hermitian settings \cite{Mos09,Ram10,Cho10,Sch10,Lon10,Cho11,Lin11,Sch13c,Vaz14,Vaz15,Lon15,Dwi16,Pil17,Kun19,Tzo21,Jin21,Mos20}.
This perspective allows us to analytically formulate the spectral conditions for a range of distinct physical phenomena, such as reflectionless scattering \cite{Ram10,Lin11}, transparency \cite{Lon15},  coherent perfect absorption \cite{Cho10,Lon10,Cho11}, and lasing \cite{Cho10,Sch10,Zha18,Par18}, and contrast these with the quantisation conditions of finite systems with open or periodic boundary conditions. From this, we can then identify the interdependence of these phenomena.

Thereby, we find that
reflectionless scattering, coherent perfect absorption, and lasing occur independently of the nonreciprocity in the system, hence, can be achieved irrespective of the extent of the skin effect, while the transparency condition involves the ensuing directed amplification explicitly, and hence can be achieved by utilizing this effect.

We further illuminate these findings in a flexible model that reveals the relevance of topological edges states for each of these physical settings. This allows us to identify three particular combined effects, where we (I) establish a direct link between the topological phase of the system and whether it can be reflectionless from one side, (II) relate invisibility to the skin-effect phase transition of the edge states, and (III) design a coherent perfect absorber that is transparent from a given side, irrespective of the topological phase.

In Sec.~\ref{sec:back} we collect the key elements of our theoretical description, which is based on a flexible tight-binding description and its corresponding transfer and scattering matrix. In Sec.~\ref{sec:conds} we classify a wide range of physical effects in terms of their boundary conditions, and describe how they depend on the non-Hermiticity and non-reciprocity of the system. The interplay of the effects is then illustrated in detail in   Sec.~\ref{sec:combeff}, for a model where they are facilitated by topological  edge states. Our conclusions are collected in Sec.~\ref{eq:concl}.
For convenience and completeness, we provide detailed Appendices on the relations between the utilized transfer and scattering matrices, as well as on the derivations of all boundary conditions.

\begin{figure}[t]
		\includegraphics[width=\columnwidth]{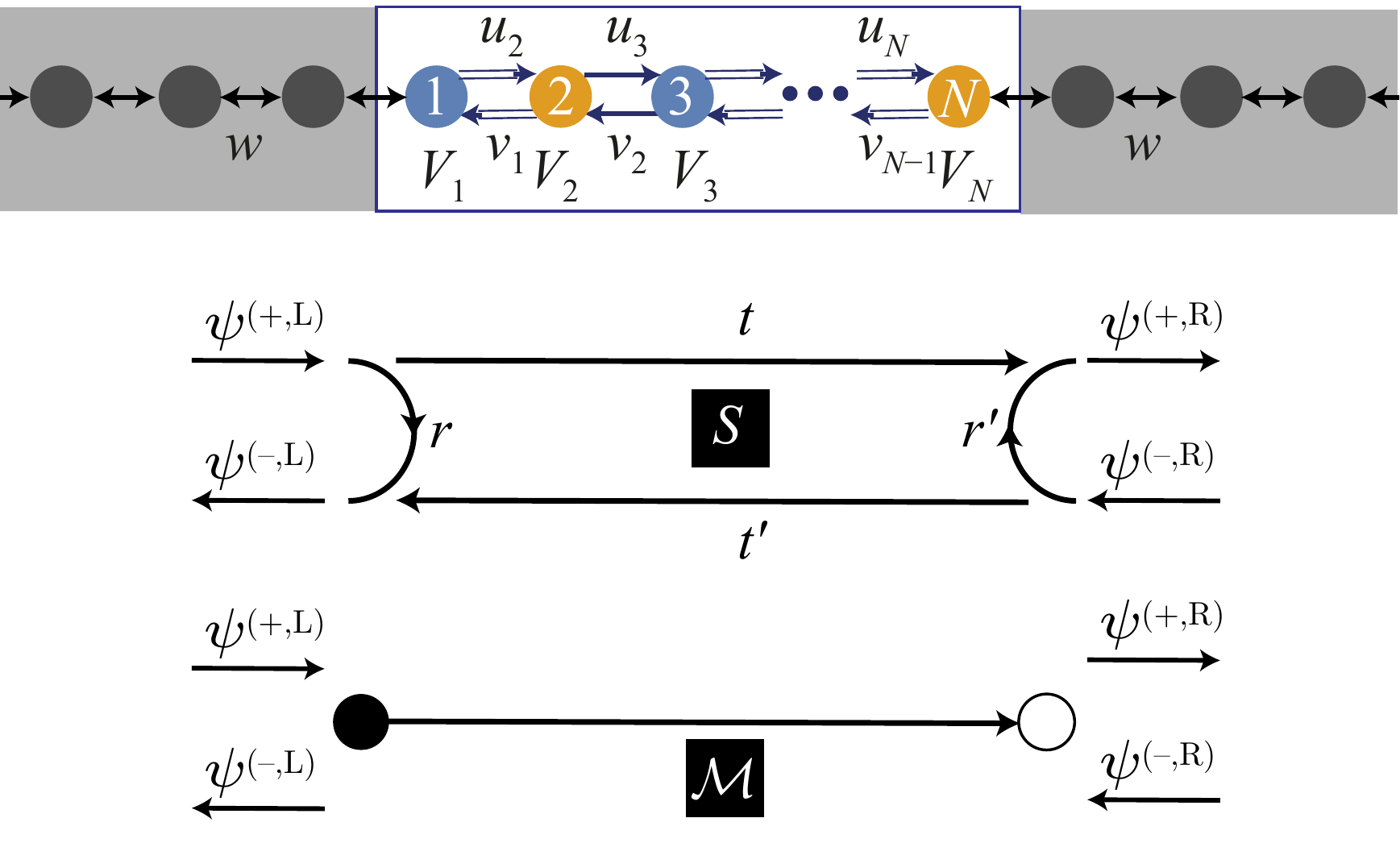}
		\caption{We study the compatibility of transport effects through non-Hermitian nonreciprocal systems, and identify by which parameters they can be adjusted.  Our theory is based on a general one-dimensional description (top), where a
chain segment (couplings $v_n\neq u_{n+1}^*$, complex onsite energies $V_n$, length $N$) is connected to featureless leads (couplings $w<0$), as described by model \eqref{eq:model}.
The elementary transport processes are captured by the scattering matrix $S$ (middle) and the transfer matrix $\mathcal{M}$ (bottom). This serves to formulate boundary conditions for a wide range of effects (see Fig.~\ref{fig1c} and Table \ref{tab1}), 
and study their interplay in general terms and concrete settings.
} \label{fig1}
\end{figure}

\section{General set-up}
\label{sec:back}
\subsection{Model}
To develop the general theory, we consider transport through a generic one-dimensional chain, as illustrated in Fig.~\ref{fig1}. This is described by a  tight-binding model
\begin{equation}
E  \psi_n=V_n \psi_n+u_n\psi_{n-1}+v_n\psi_{n+1},
\label{eq:model}
\end{equation}
where $V_n$ are onsite potentials,
$u_n$ are nearest-neighbour couplings from left to right, and $v_n$ are nearest-neighbour couplings from right to left. 
These parameters describe features of discrete components of the system, which could constitute resonators or waveguides in the optical case, while in other settings they may represent, for instance, electronic or robotic elements \cite{Bra19,Gha20,Hof20,Hel20}.
As we are interested in the case of  nonreciprocal non-Hermitian transport, we allow for situations where at least some of the couplings obey $u_{n+1}\neq v_n^*, v_n$ \footnote{If $u_{n+1}= v_n\neq v_n^*$, the system is non-Hermitian but reciprocal; see for instance Ref.~\cite{Gol14}.},
and also allow the onsite potentials $V_n$ to be complex.
The system is confined to the region $1\leq n\leq N$, while the remaining sites describe the leads.
We model these leads in the featureless wide-band limit,
which is obtained from constant
couplings $u_n=v_n=w<0$ ($n\leq 0$, left lead, or $n\geq N$, right lead) with the potential energy  tuned to the band centre ($V_n=E$).
The propagating waves then have the simple form
\begin{align}
&\psi_n=\psi^{(+)}i^{n-n_+} &\quad\mbox{(propagating to the right)},\nonumber\\
&\psi_n=\psi^{(-)}(-i)^{n-n_-} &\quad\mbox{(propagating to the left)},\label{eq:leadwaves}
\end{align}
where the amplitudes $\psi^{(\pm)}$ are position-independent throughout a given lead. The possibly non-integer offsets $n_\pm$  can  be chosen separately in each lead, and account for the $U(1)$ gauge freedom.
We assume that the boundary couplings from the leads
to the system match perfectly, $u_1=v_{N}\equiv w$, which does not imply any restrictions as one can always include the first site of the lead into the system.
Thereby the only parameter characterizing the leads is $w$, which controls the transparency of the contacts.


\subsection{Transport framework}
To characterize the system from a transport perspective, our main building blocks are the one-step real-space transfer matrix
\begin{align}
M_{n}=\left(\begin{array}{cc}
(E -V_n)/v_n & -u_n/v_n\\
 1 & 0
 \end{array}
\right),
\end{align}
the resulting real-space transfer matrix
\begin{align}
M=M_{N} \cdots M_3M_2M_1
\end{align}
of the complete system,
the corresponding transfer matrix
\begin{align}
 \mathcal{M}&=\frac{1}{2}\left(\begin{array}{cc}-i & 1 \\ 1 & -i \end{array}\right)
M\left(\begin{array}{cc}i & 1 \\ 1 & i \end{array}\right)
\label{eq:realtopropmaintext}
\end{align}
in the propagating-state basis,
and, finally, the scattering matrix
\begin{align}
S&\equiv\left(\begin{array}{cc}r & t'\\ t &  r'\end{array}\right)
=\frac{1}{\mathcal{M}_{22}}
\left(\begin{array}{cc}-\mathcal{M}_{21}& 1\\
\det\mathcal{M}
 &  \mathcal{M}_{12}\end{array}\right).
 \label{eq:smatmaintext}
\end{align}
The chain of expressions relates the scattering matrix to the underlying model \eqref{eq:model}. Furthermore, this relation can  also be expressed more directly as
\begin{align}
S
&=
-i\openone+2w(E -H_\mathrm{eff})^{-1}_{\{1,N\}},
\label{eq:mw2maintext}
\\
H_\mathrm{eff}&=H+w\,\mathrm{diag}\,(i,0,\ldots,0,i),
\label{eq:heffmain}
\end{align}
where the indices indicate a $2\times 2$ matrix formed of the corner elements of an $N\times N$ matrix, which here represents the Greens function of the open system with an effective Hamiltonian $H_\mathrm{eff}$ that includes the self energy of the leads (see  App.~\ref{app:a} for further details).
These matrices capture the transport features in terms of linear relations between the propagating wave amplitudes in the leads, as illustrated in Fig.~\ref{fig1}.

\begin{figure}[t]
		\includegraphics[width=\columnwidth]{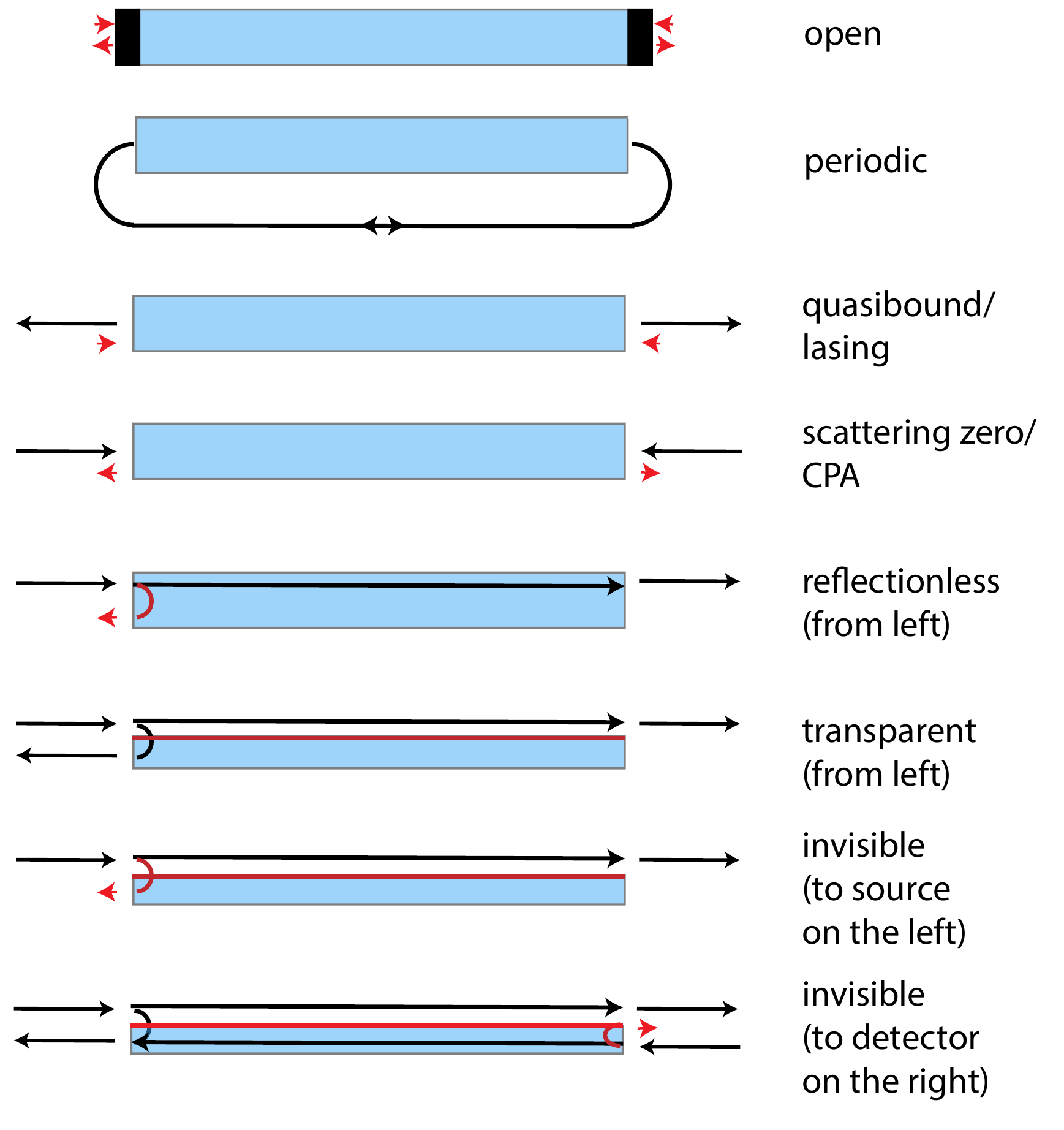}
		\caption{Illustration of transport effects and their boundary conditions. The red arrows indicate forbidden processes, while the red line in the transparent system highlights that there, the transmitted signal is identical to that obtained bypassing the system.
We study the relations between these effects, identify the parameters they depend on, and describe how they can be combined.} \label{fig1c}
\end{figure}

\begin{table*}[th]
\caption{Comparison of boundary conditions (BCs) and energy constraints in different physical situations.
In all cases, the transfer matrix 
$\mathcal{M}$ and scattering matrix $S$ have to  be taken as  functions of  energy $E$. 
When energies are stipulated as real this is to guarantee stationary situations, while complex energies refer to quasi-stationary behavior. 
}
\label{tab1}
\begin{tabular}{cccc}
  \hline
  situation & transfer BC & scattering  BC & energy constraints\\
  \hline
  transport &
$\left(\begin{smallmatrix}\psi^+_R\\ \psi^-_R\end{smallmatrix}\right)
=\mathcal{M}\left(\begin{smallmatrix}\psi^+_L\\ \psi^-_L\end{smallmatrix}\right)$
&
$
\left(\begin{smallmatrix}\psi^-_L\\ \psi^+_R\end{smallmatrix}\right)
=S\left(\begin{smallmatrix}\psi^+_L\\ \psi^-_R\end{smallmatrix}\right)
$
& real, given
\\[.2cm]
  open
  &
  $(\begin{smallmatrix}i & 1\end{smallmatrix})
  \mathcal{M}
   \left(\begin{smallmatrix}-i \\ 1\end{smallmatrix}\right)=0$
&
$\det (S-i\openone)=0$
& discrete, complex
 \\[.2cm]
  periodic
   & $\det(\mathcal{M}-e^{ik}\openone)=0$
    & $\det[ S-\sigma_x\,\mathrm{diag}\,(e^{ik},e^{-ik})]=0$
    & complex function of real $k$
    \\[.2cm]
  quasi-bound
 & $\mathcal{M}_{22}=0$
& $\det S^{-1}=0$
& discrete, complex\\[.2cm]
  lasing
 & $\mathcal{M}_{22}=0$
& $\det S^{-1}=0$
& discrete, real\\[.2cm]
  scattering zero
  & $\mathcal{M}_{11}=0$
 &$\det S=0$
  & discrete, complex
  \\[.2cm]
  CPA
  & $\mathcal{M}_{11}=0$
 &$\det S=0$
  & discrete, real
    \\[.2cm]
 reflectionless
    &$\mathcal{M}_{21}=0$ or $\mathcal{M}_{12}=0$& $r=0$ or $r'=0$
    &real
    \\[.2cm]
 transparent
    &
    $(\mathcal{M}^{-1})_{11}\mbox{ or }\mathcal{M}_{22}=(-i)^N$
    & $t=i^N$ or $t'=i^N$
    &real
     \\[.2cm]
  \hline
\end{tabular}
\end{table*}

\section{Classification of transport effects}
\label{sec:conds}
\subsection{Boundary conditions}
Equipped with these expressions, we can formulate, as our first main goal, a comprehensive set of boundary conditions for a range of physical effects, as summarised in Table \ref{tab1} (see Fig.~\ref{fig1c} for illustration of the physical effects, and App.~\ref{app:b} for detailed derivations).
In these conditions, the transfer matrix
$\mathcal{M}$ and scattering matrix $S$ have to be taken as  functions of  energy $E$, which links transport and spectral features.

In \emph{stationary transport} settings, the energy is real and given, while in other situations the conditions have to be read as implicit equations, and typically lead to a discrete complex spectrum describing quasi-stationary behaviour. Examples of stationary behaviour are systems that are \emph{reflectionless} or \emph{transparent} when probed from one side (the stated condition is for strict transparency, both for the intensity as well as the phase of the transmitted signal).

A system that is both reflectionless and transparent from certain sides is \emph{invisible} with respect to a suitably placed source (if the sides are the same) or detector (if they are opposite).

Complex energies describe quasi-stationary behavior, where solutions with $\mathrm{Im}\,E<0$ describe resonant modes with a finite life time, while for $\mathrm{Im}\,E>0$ the modes display a transient exponential growth that physically can only be sustained until nonlinear saturation effects set in.

For \emph{quasi-bound states}, the corresponding  quasi-stationary wavefunctions fulfill purely outgoing conditions. If this is achieved at a real energy, the system serves as a stationary emitter of coherent radiation, as encountered in a \emph{laser}.
In both cases, the energies are determined by the poles of the scattering matrix, which by Eq.~\eqref{eq:heffmain} furthermore coincide with the eigenvalues of the effective Hamiltonian. 

Interchanging the role of incoming and outgoing states, we arrive at the spectrum of \emph{scattering zeros}, which when real allow the system to realize \emph{coherent perfect absorption} (CPA).

The Table also contains entries for finite \emph{open} and \emph{periodic} systems, where the leads are fictitious elements in the construction of the conditions. The quantization condition for a finite open system can be derived in the limit of quasibound states with pinched-off leads, $w\to 0$, and
corresponds to vanishing amplitudes on the first site in each fictitious lead. For periodic systems, energies become parameterized by the Bloch wave number $k$.

\subsection{Conditions for nonreciprocal and non-Hermitian transport}
\label{sec:nonrecnonher}

Our second main goal is to investigate how these conditions of various specific physical transport effects relate to the
general physical features of non-Hermiticity and nonreciprocity. We base this on the following general definitions, which at the same time help to quantify these effects.

Nonreciprocal transport is defined by
\begin{equation}
S\neq S^T,
\end{equation}
so that $t\neq t'$.
For the transfer matrices, this implies
\begin{align}\label{eq:mnonrec}
D\equiv\det M=\det \mathcal{M}=\prod_n (u_n /v_n)\neq 1.
\end{align}
Furthermore, in a non-Hermitian system the scattering matrix is no longer unitary,
\begin{equation}
SS^\dagger \neq \openone,
\end{equation}
which implies
\begin{align}\label{eq:mnonherm}
M^\dagger \sigma_y M\neq \sigma_y,
\\
 \mathcal{M}^\dagger \sigma_z\mathcal{M}\neq \sigma_z.
\end{align}

In the derivations of the boundary conditions in Tab.~\ref{tab1}, we took care that they do not imply any of these relations. Therefore, equipped with these additional definitions, we can now highlight whether they provide any further constraints, or indeed matter at all.

To do so, we define the reciprocal counterpart of a system by setting its couplings to $\bar v_n=\bar u_{n+1}=\sqrt{v_nu_{n+1}}$. Accounting also for the boundary conditions at the leads, the corresponding reciprocal transfer matrix then becomes expressed as $\overline{M}=M/\sqrt{D}$,  $\overline{\mathcal{M}}=\mathcal{M}/\sqrt{D}$.
It then becomes apparent that all conditions from the Table where transfer matrix elements have to vanish are identical in both variants of the system. 

The significance of this observation is further clarified when we consider the system to be periodic, where we allow the unit cell to have arbitrary length. Setting the transfer matrix of the unit cell to $M_c$ and $\overline{M}_c=M_c/\sqrt{d}$ where $d=\det M_c$, the transfer matrix of the system with $L$ unit cells can be written as
\begin{equation}
M(E)=d^{L/2}[U_{L-1}(z)\overline{M}_{\mathrm{c}}-U_{L-2}(z)\openone]
,
\label{eq:mallexact}
\end{equation}
where $U_l(z) $ are the Chebyshev polynomials of the second kind and
$
z=\mathrm{tr}\,\overline{M}_{\mathrm{c}}/2
$.
This clearly separates out the effect of directed amplification, which scales the transport from left to right by a factor $d^{L/2}$, while in the opposite direction it is scaled by $d^{-L/2}$. In contrast, all characteristics that rely on vanishing matrix elements are the same in the nonreciprocal and reciprocal variant of the system. 

 In practice, this implies that when a system is tuned to exhibit effects such as reflectionless transport, coherent perfect absorption, or lasing, its amount of directed amplification can still be independently modified. On the other hand, given that $\det\mathcal{M}$ is finite, it is not possible to make a laser or coherent perfect absorber reflectionless from any side. We give practical examples of compatible combinations in the next section, where we discuss a topological model system.

\section{Application and interplay of effects}
\label{sec:combeff}
\subsection{Illustrative model system}
As our third main goal, we illustrate our general statements for a complex nonreciprocal dimer chain, defined by  alternating complex onsite potentials
\begin{subequations}
\begin{align}
V_{2l-1}=i\gamma,\quad V_{2l}=i\gamma'
, \label{eq:par1}
\end{align}
and asymmetric alternating couplings
\begin{align}
u_{2l-1}=u, \quad u_{2l}=u',\quad  v_{2l-1}=v, \quad v_{2l}=v',
\label{eq:par2}
\end{align}
\label{eq:par}%
\end{subequations}
where $l=1,2,\ldots,L$ enumerates the unit cells of a system with $N=2L$ sites.
Keeping all parameters real, this model encompasses a range of special cases exhibiting different symmetries of topological significance, with the Hermitian limit $u=v', v=u'$, $\gamma=\gamma'=0$ defining the Su-Schrieffer-Heeger (SSH) model \cite{Su79}, non-Hermitian cases with complex scalar potentials including PT-symmetric and charge-conjugation-symmetric \cite{Rud09,Ram12b,Sch13,Zhu14,Pol15,Zeu15,Jin17,Lie18,Lan18,Zha18,Par18,Pan18,Mos20a} systems, and non-reciprocal variants with imaginary vector potentials encompassing  those at the heart of the study of the non-Hermitian skin effect \cite{Yao18,Xio18,Kun18,Mar18,Yao18b,Lon19,Yok19,Son19,Lee19,Jin19,Oku20,Yi20,Yuc20,Bra19,Gha20,Hof20,Hel20,Sch20}.
For our discussion, we employ a nonunitary similarity transformation within each unit cell to set $v'=u=w$, with the latter equality corresponding to ballistic coupling to the leads. Given that mirror-reflecting the system  corresponds to the transformation $(u,u',\gamma)\leftrightarrow (v',v,\gamma')$, we furthermore assume $\gamma\geq \gamma'$.
The key parameters of the system then are
\begin{equation}
d=\frac{u'}{v}, \quad \kappa=\frac{v u'}{w^2},
\label{eq:param}
\end{equation}
where $d$ quantifies the amount of nonreciprocity and $\kappa$ captures the topological characteristics inherited from the SSH limit.

\begin{figure}[t]
		\includegraphics[width=\columnwidth]{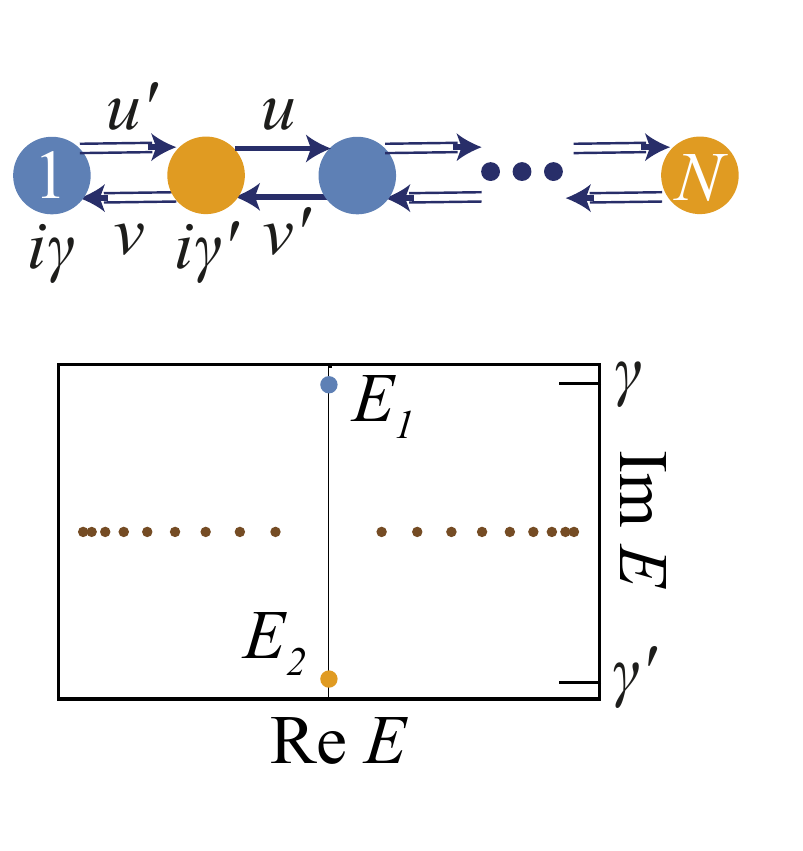}
		\caption{Complex energy spectrum of a finite system with closed boundary conditions, based on model \eqref{eq:model} with parameters as defined in Eqs. \eqref{eq:par} and \eqref{eq:param}. The two states with purely imaginary energies $E_1$ and $E_2$ arise from the topological edges states of the system, which exist for dimerization parameter $\kappa<1$, and are well isolated from other states on the imaginary axis as long as the gain-loss contrast $|\gamma-\gamma'|<|2w(1-\kappa)|$.
 In the figure,   $\kappa=0.4$ and $\gamma-\gamma'=w(\kappa-1)$. Notably, this  spectrum is independent of the non-reciprocity parameter $d$, which, however, determines the edge where these states are localized at (see phase diagram in Fig.~\ref{fig:phasediag}) .
} \label{fig2}
\end{figure}

\begin{figure}[t]
	\begin{center}
		\includegraphics[width=\columnwidth]{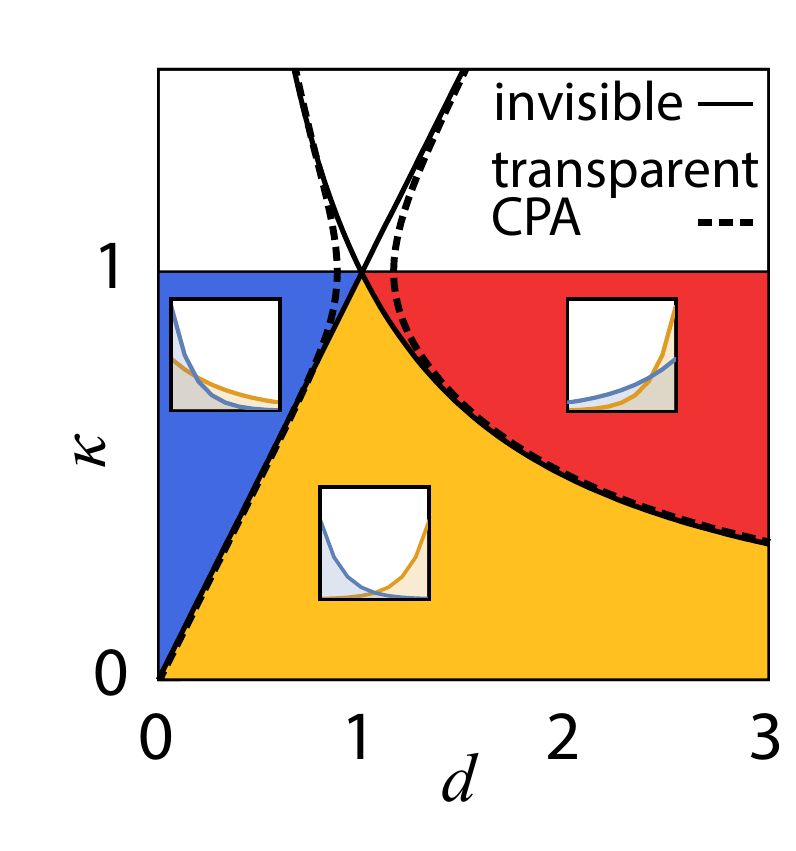}
		\caption{Relation of transport effects to the skin effect of topological zero modes in the model of Fig.~\ref{fig2}. The topological phase $\kappa<1$, where the finite system displays edge states, is further broken down into three phases with transitions at nonreciprocity parameter $d=\kappa, 1/\kappa$. These denote the critical parameter where the edge states relocalize from one edge to the other by means of the non-Hermitian skin effect, as illustrated in the inset. In the central phase, the edge states are localized at opposite edges, as is also the case in the Hermitian limit. In the other two phases, both states are localized at the same edge. The superimposed curves highlight the relation to combined transport effects in the system connected to leads, as further detailed in the text. At the skin-effect phase transition and its extrapolation (thin solid curves) the edge state can be made invisible to either a source or detector. The thick curves indicate conditions where one can combine transparency with  coherent perfect absorption in a system of length $L=10$.} \label{fig:phasediag}
	\end{center}
\end{figure}


As shown in the phase diagram in Fig.~\ref{fig:phasediag},
topological edge states exist for $\kappa<1$, and then take the form of spectrally isolated, sublattice-polarized zero modes with energies pinned to $E_{1}=i\gamma$ and $E_{2}=i\gamma'$, hence $\mathrm{Re}\,E=0$.
For
$\kappa d<1$, the state with energy $E_1$ is localized at the left edge, while for
$\kappa d>1$  it is localized at the right edge; the latter situation can only occur in the nonreciprocal system.
The state with energy $E_2$ is localized at the right edge when
$\kappa<d$, and localized at the left edge when $\kappa>d$, where the latter case is again only attainable in the nonreciprocal system.

These relocalization phenomena are a manifestation of the skin effect for the topological modes, and can be made physically visible, e.g., by probing the finite system via external driving \cite{Sch20}. 
The stated conditions apply to the mathematically-spoken `right' eigenvectors. External driving also reveals the role of the
biorthogonal `left' eigenvectors, which here are obtained by the transformation $(u,u')\leftrightarrow (v',v)$ while keeping $\gamma$ and $\gamma'$ unchanged.
The biorthogonal eigenstate with energy $E_1$ is therefore localized at the left edge if $\kappa<d$ and at the right edge if $\kappa>d$; the biorthogonal eigenstate with energy $E_2$ is localized at the right edge if $\kappa d<1$ and at the left edge if $\kappa d>1$.

We find that additional bulk zero modes with $\mathrm{Re}\,E=0$ appear if the gain-loss contrast $(\gamma-\gamma')^2/w^2>4(1-\sqrt{\kappa})^2$; furthermore, bulk states become dynamically unstable if $\kappa$ (and hence also $d$) is negative, where the mapping to Hermitian couplings breaks down. In the remainder we focus on the situation with two clearly defined edge states, so that we can directly examine their relevance and influence on the physical transport characteristics of the system.

Given our assumption $\gamma\geq \gamma'$, the edge mode with energy $E_1$ is then the most stable mode in the system, once it exists, and the other edge mode is the least stable mode in the system.  Overall, the system is then dynamically stable for $\mathrm{Im}\,E\geq \gamma$, and so effectively we are permitted to set the energy to $E=E_1$, which we will do as soon as we have established the conditions for each physical scenario.

While the system can be easily studied numerically, it can also be conveniently  analysed based on the
 exact expression  \eqref{eq:mallexact} for the transfer matrix of the whole system,
where now
\begin{align}
M_{\mathrm{c}}
=\sqrt{\frac{d}{\kappa}}\left(\begin{array}{cc}
	\frac{(E -i\gamma)(E -i\gamma')}{w^2}-\kappa&-\frac{(E -i\gamma')}{w}  \\
	\\
	\frac{(E -i\gamma)}{w}&-1
	\end{array}\right)
\label{transfer matrix for non-Hermitian SS model}
,
\end{align}
$z=d^{-1/2} \mathrm{tr}\, M_c/2$, and $d$ and $\kappa$ as defined in Eq.~\ref{eq:param}.

\subsection{Effect I: Reflectionless transport via edge states and the role of the topological phase}
We start with reflectionless transport, where we establish a link to the topological phase of the system.
With the help of Eq.~\eqref{transfer matrix for non-Hermitian SS model}, the corresponding  boundary condition from the Table can be rephrased as
\begin{equation}
\mathcal{F}U_{L-1}(z)=0
\label{implicit2:invisibility I}
\end{equation}
with
\begin{equation}
\mathcal{F}=(E-i\gamma)(E-i\gamma')+w^2(1-\kappa)\mp(\gamma-\gamma')w,
\end{equation}
where the upper (lower) sign applies to probing the system from the left (right).
Given that condition \eqref{implicit2:invisibility I} factorises, we find two different types of solutions, which we can naturally interpret as a \emph{local} and a \emph{global} mechanism for reflectionless transport. The global mechanism arises from the Chebyshev nodes, $U_{L-1}(z)=0$, and hence depends on the length of the system, but is independent on the side from which one probes the system. Indeed, we find that this global mechanism is essentially independent of the existence of edge states: at $E=E_1$ and $E=E_2$, $z$ is invariant under the transformation $\kappa\to 1/\kappa$, which connects the topological phase ($\kappa<1$) and nontopological phase ($\kappa>1$) where edge states do or do not exist at these energies.
This is in contrast to the local condition, $\mathcal{F}=0$, where $\kappa$ enters as an essential parameter. For illustration, let us again set $E=E_1$, hence, to the energy of the most stable edge state when it exists, and consider to probe the system from the \emph{left}. We then find
that the system is reflectionless when the gain/loss contrast takes the specific value
\begin{equation}
\gamma-\gamma'=w(1-\kappa).
\label{eq:refleft}
\end{equation}
Recalling that we formulated all conditions for $w<0$, the right-hand side is negative in the topological phase, meaning that this would require $\gamma<\gamma'$, in contradiction with our assumption; therefore the system can only  be made reflectionless from this side in the nontopological phase.
However, the system  can indeed be made reflectionless in the topological phase when we probe it from the \emph{right}, where we require the opposite gain contrast
\begin{equation}
\gamma-\gamma'=-w(1-\kappa).
\label{eq:refright}
\end{equation}
Thus, the conditions for reflectionless transport leave a clear physical signature of the topological phase we operate in.

In all these reflectionless scenarios, the roles of left and right are interchanged when we instead assume $\gamma<\gamma'$. Therefore, in more general terms, and taking the Hermitian limit as the reference point for the localization of these states, this means that reflectionless transport can be achieved in the topological phase when we place the more instable of the two edge state at the near end of the system, and the more stable one at the far end.

\subsection{Effect II: Invisibility at the skin-effect phase transition}
As  condition \eqref{eq:refright} is independent of the nonreciprocity parameter $d$, the reflectionless transport from the right is maintained even when any of the edge states relocalises via the skin effect to the other edge of the system. This parameter can therefore be tuned to achieve other compatible transport effects. In particular, keeping the system in the topological phase, and reflectionless from the right according to Eq.~\eqref{eq:refright}, we can make it strictly transparent from the right (left) by setting $d=\kappa$  ($d=1/\kappa$), so that the system is then invisible to a source (detector) placed to the right of the system. These conditions of $\kappa$ and $d$ coincide exactly with the skin-effect phase transition of the topological states, which thereby form naturally transparent channels exactly at this transition. This uncovers a direct transport signature of the skin effect.

\subsection{Effect III: Transparent coherent perfect absorber}
As stated above, the parameter $d$ also drops out of other transport conditions, such as for lasing and for coherent perfect absorption. These conditions can again be made explicit by utilizing Eq.~\eqref{transfer matrix for non-Hermitian SS model}. For  coherent perfect absorption, this results in the following equation for the required gain contrast,
\begin{equation}
\gamma-\gamma'=-w\frac{(\kappa-1)(\kappa^L+1)}{\kappa^L-1}.
\label{eq:cpa}
\end{equation}
This is always positive (recalling again $w<0$), independent of $d$, and invariant under the replacement $\kappa\to 1/\kappa$, so that stable CPA conditions can be achieved irrespective of the topological phase or the extent of the directed amplification. With this CPA condition in place, we can still make the system simultaneously transparent from the left by setting
$d=(\kappa^{L/2} +\kappa^{-L/2})^{2/L}$, while it is transparent from the right for $d=(\kappa^{L/2} +\kappa^{-L/2})^{-2/L}$, demonstrating also for this combination that compatible effects can indeed be tuned independently in a specific model.
In Fig.~\ref{fig:phasediag}, the thick curve illustrates this conditions for $L=10$. Notably, for $L\to\infty$, the condition coincides again with the skin-effect phase transition.

\section{Conclusions}
\label{eq:concl}

In summary, we established relations and distinctions between a range of discrete transport effects in non-Hermitian, nonreciprocal and potential topological systems, which we characterised from a unifying scattering perspective. This allowed us to identify effects that are compatible and independent of each other, hence, can be achieved simultaneously by tuning suitable parameters. The parameter determining directed amplification due to non-Hermitian nonreciprocity plays a distinguished role as it modifies the transparency of the system independently from the conditions for a wide range of other effects. In a concrete model system we showed how these signatures are further linked to other characteristics, such as the existence of topological edge states. As we showed for three effects,
this perspective can be usefully applied to concrete models, and then utilised to design devices that combine specific characteristics.

The provided framework should prove useful as starting point for further investigations, including specific systems, related transport effects, or extended settings. In particular, while we illustrated here that these effects can be achieved already in one-dimensional tight-binding models,
we note that this approach can be extended to quasi-one dimensional systems, allowing also to explore topological models in higher dimensions \cite{Kun19}, and further be enriched by considering symmetry constraints on the scattering description \cite{Jin21}. For the  Reader who would like to apply or transfer our insights to specific settings, the key results are in Table \ref{tab1}, where we collect the boundary conditions for the studied range of effects (illustrated in Fig.~\ref{fig2}), and
Sec.~\ref{sec:nonrecnonher}, where we discuss their relation to precisely defined transport notions of non-Hermiticity and nonreciprocity. For a key illustration of the resulting physical interplay of the effects with topological states, we refer to the phase diagram of Fig.~\ref{fig:phasediag}. In particular, experimentalists should feel encouraged to consider if the combined effects in Sec.~\ref{sec:combeff} can be realized on their platform.

\begin{acknowledgments}
The authors acknowledge funding by EPSRC via Programme Grant No. EP/N031776/1.
\end{acknowledgments}

\appendix

\section{Transfer scattering and  matrices}
\label{app:a}
Here we collect the specific definitions and relations between transfer and scattering matrices for the case of the featureless leads defined in the main text, allowing to assure that we adopt expressions that remain valid for systems with complex scalar and vector potentials.

\subsection{Real-space transfer matrix}
We start with the conventional one-step real-space transfer matrix, defined through the relation
\begin{align}
\left(\begin{array}{c}\psi_{n+1}\\ \psi_n\end{array}\right)
=M_{n}\left(\begin{array}{c}\psi_{n}\\ \psi_{n-1}\end{array}\right),
\end{align}
which for model \eqref{eq:model}
of the main text
implies
\begin{align}
M_{n}=\left(\begin{array}{cc}
(E -V_n)/v_n & -u_n/v_n\\
 1 & 0
 \end{array}
\right).
\end{align}
For the complete segment,
the real-space transfer matrix is defined as
\begin{align}
\left(\begin{array}{c}\psi_{N+1}\\ \psi_N\end{array}\right)
=M\left(\begin{array}{c}\psi_{1}\\ \psi_{0}\end{array}\right),
\label{eq:Mtot}
\end{align}
giving
\begin{align}
M=M_{N} \cdots M_3M_2M_1.
\end{align}

Equation \eqref{eq:Mtot} not only involves the sites at the end of the system, but also the first sites in the leads.  This facilitates the formulation of a wide range of boundary conditions, including for closed systems where the sites are fictitious.

\subsection{Propagating-state transfer matrix}

We next formulate the transfer matrix in the propagating-wave basis, denoted by  $\mathcal{M}$.
This transfer matrix can then be used to formulate the general
scattering  boundary conditions that are central to the description of transport,
which is followed by a discussion of the relation to other boundary conditions.

As in the main text, we adopt feature-less leads in the wide-band limit and assuming uniform couplings $w<0$.
The sign of $w$ can  be chosen freely by exploiting the  $Z_2$ gauge freedom  $\psi'_n= (-1)^n\psi_n$. While this
changes the sign of the group velocity, and thereby
reverts the propagation directions, this freedom leaves the transport characteristics invariant.
Tuning the leads to the band center
ensures that the self-energy of the leads is energy-independent, so that the leads are indeed featureless, and using identical couplings in both leads gives
propagating waves that carry the same flux, so that further flux normalization is not required.

In the propagating-state basis of Eq.~\eqref{eq:leadwaves}
of the main text,
the transfer matrix is then defined by the relation
\begin{equation}
\left(\begin{array}{c}\psi^{(+,R)}\\ \psi^{(-,R)}\end{array}\right)
=\mathcal{M}\left(\begin{array}{c}\psi^{(+,L)}\\ \psi^{(-,L)}\end{array}\right),
\label{eq:proptrans}
\end{equation}
where $L$ and $R$ refer to the left and right lead.
For this, we invoke the wave-matching condition of propagating waves in real space,
\begin{align}
\left(\begin{array}{c}i\psi^{(+,R)}+\psi^{(-,R)}\\ \psi^{(+,R)}+i\psi^{(-,R)}\end{array}\right)
=M\left(\begin{array}{c}i\psi^{(+,L)}+\psi^{(-,L)}\\ \psi^{(+,L)}+i\psi^{(-,L)}\end{array}\right),
\label{eq:propcondition}
\end{align}
which is obtained using convenient offsets in Eq.~\eqref{eq:leadwaves}.
Using
\begin{align}
\left(\begin{array}{c}i\psi^{(+)}+\psi^{(-)}\\ \psi^{(+)}+i\psi^{(-)}\end{array}\right)
=
\left(\begin{array}{cc}i & 1 \\ 1 & i \end{array}\right)
\left(\begin{array}{c}\psi^{(+)}\\ i\psi^{(-)}\end{array}\right),
\end{align}
we then obtain
\begin{align}
 \mathcal{M}&=\frac{1}{2}\left(\begin{array}{cc}-i & 1 \\ 1 & -i \end{array}\right)
M\left(\begin{array}{cc}i & 1 \\ 1 & i \end{array}\right)
,
\label{eq:realtoprop}
\end{align}
 where the matrix elements follow the pattern
\begin{align}
\mathcal{M}_{ab}=\frac{1}{2}(M_{ab}+M_{\bar a\bar b}+iM_{\bar a b}-i M_{a\bar b}),
\label{eq:realtopropelements}
\end{align}
with $\bar 1=2$ and $\bar 2=1$.

To illustrate the consistency of the featureless wide-band limit, note that such the leads  are themselves described by real-space transfer matrices
$M^{(\mathrm{lead})}=-i\sigma_y$, so that translating the propagating states by one site amounts to
\begin{align}
\mathcal{M}^{(\mathrm{lead})}=i\sigma_z.
\end{align}
This indeed corresponds to the phase factors picked up by the propagating waves according to
Eq.~\eqref{eq:leadwaves}.

\subsection{Scattering matrix}

Given the described transformation to propagating waves in featureless leads,
scattering boundary conditions can now be implemented as in a space-continuous system, where they are captured by the scattering matrix
\begin{align}
\left(\begin{array}{c}\psi^{(-,L)}\\ \psi^{(+,R)}\end{array}\right)
=S\left(\begin{array}{c}\psi^{(+,L)}\\ \psi^{(-,R)}\end{array}\right).
 \label{eq:smat2}
\end{align}
Here $r$ and $t$ are the reflection and transmission amplitudes for an incoming wave from the left lead, whilst $r'$ and $t'$ are the corresponding amplitudes for the right lead.
(Note that we here exploit that the propagating states in both leads carry the same flux, and  that the couplings from the system to the leads are the same. If the couplings $w_L$ and $w_R$ in both leads differ, but the wide-band limit remains applied, the scattering amplitudes have to be scaled by factors $\sqrt{|w_{l,R}|}$ to reflect the different group velocities.)

To relate the scattering matrix to the transfer matrix,
we demand
\begin{align}
&
\left(\begin{array}{c}t \\  0 \end{array}\right)
=\mathcal{M}\left(\begin{array}{c}1\\ r\end{array}\right),
\quad\left(\begin{array}{c}r'\\1 \end{array}\right)
=\mathcal{M}\left(\begin{array}{c} 0 \\ t' \end{array}\right),
\end{align}
which gives
\begin{align}
S&\equiv\left(\begin{array}{cc}r & t'\\ t &  r'\end{array}\right)
=\frac{1}{\mathcal{M}_{22}}
\left(\begin{array}{cc}-\mathcal{M}_{21}& 1\\
\det\mathcal{M}
 &  \mathcal{M}_{12}\end{array}\right).
 \label{eq:smat}
\end{align}
The apparently asymmetric form of these relations arises from the sense of direction embodied in the transfer matrix. That the physical symmetry is fully respected follows when we straightforwardly
rewrite the components in terms of the inverted transfer matrix, such as $t=1/(\mathcal{M}^{-1})_{11}$.

Equation \eqref{eq:smat} can be inverted
to give
\begin{align}
\mathcal{M}=\frac{1}{t'}
\left(\begin{array}{cc}-\det S& r'\\
-r
 &  1 \end{array}\right).
\end{align}
Furthermore, using the rule \eqref{eq:realtopropelements}, the scattering amplitudes can  be expressed directly in terms of the real-space transfer matrix or vice versa, where one can  conveniently employ $\det\mathcal{M}=\det M$.

These expressions relate the scattering matrix to the underlying model \eqref{eq:model}. However, exploiting the composition rules of scattering matrix from different segments, this relation can also be expressed more directly as
\begin{align}
S&=-i\frac{1+iw(E -H)^{-1}_{\{1,N\}}}{1-iw(E -H)^{-1}_{\{1,N\}}}
,
\label{eq:mw1}
\end{align}
where the indices indicate a $2\times 2$ matrix formed of the corner elements of an $N\times N$ matrix, which here represents the Greens function of the closed system.
Resummation of the corresponding power series then results in
Eq.~\eqref{eq:mw2maintext}
 from the main text,
where  we now encounter the Greens function of the open system with an effective Hamiltonian $H_\mathrm{eff}$ that includes the self energy of the leads.

\section{Detailed formulation of boundary conditions}
\label{app:b}

Equation~\eqref{eq:smat2} encompasses very general scattering boundary conditions, which can be further specified depending on the nature of the source, such as for scattering from the left lead. Here, we describe in detail how this can be employed to arrive at the boundary conditions collected in Table~\ref{tab1}
of the main text, including for open and periodic systems where the leads are fictitious.

\subsection{Open boundary conditions}

For a closed system with open boundary conditions,
the bound states are the eigenvalues of the Hamiltonian $H$ of the truncated system.
In our setting, these boundary conditions can be implemented by setting
 $\psi_0=\psi_{N+1}=0$ on the first sites in the leads, which then are fictitious. In terms of the real-space transfer matrix, these boundary conditions correspond to
\begin{align}
M_{11}(E )=0,\label{implicit:open}
\end{align}
which has to be read as an implicit equation for the bound-state energies $E_n$, where
the solutions are in general discrete, but possibly complex.
One can check that as required, this form of the quantization condition is still independent of the values $w$ for the couplings to the then fictitious leads.
In terms of the propagating waves, this condition reads
\begin{align}
\mathcal{M}_{11}(E )+\mathcal{M}_{22}(E )-i\mathcal{M}_{21}(E )+i\mathcal{M}_{12}(E )=0,
\end{align}
which  in Table \ref{tab1}
we have written in a more compact form.
In terms of the transport coefficients we can formulate this as a standard scattering quantization condition,
\begin{align}
\det (S(E )-i\openone)=0,
\label{eq:scondopen}
\end{align}
which in both cases again are  implicit equations for the bound-state energy.
From Eq.~\eqref{eq:mw1}, we furthermore see that condition \eqref{eq:scondopen} is equivalent to finding the poles of the resolvent $[E -H]^{-1}_{\{1,N\}}$, which are indeed the eigenvalues of $H$.

\subsection{Periodic boundary conditions}

In a periodic system we require $\psi_{n+N}=e^{ik} \psi_{n}$
where $k$ is real. In terms of the eigenvalues $\lambda_l$ ($l=1,2$) of the matrix $M$ , this amounts to the implicit equations
\begin{align}
|\lambda_{1}(E )|=1 \mbox{ or } |\lambda_{2}(E )|=1
\label{eq:periodiccond}
\end{align}
for the energy $E $.
(At degeneracies, including exceptional points, we set $\lambda_1(E)=\lambda_2(E)$, reflecting the algebraic multiplicity of the eigenvalues, but not necessarily their geometric multiplicity.)
In general, the solutions form curve segments in the complex plane, where the phase $k$ varies along the segment.
Therefore, we can interpret Eq.~\eqref{eq:periodiccond} as a condition
for the dispersion relation $E (k)$, which is more directly obtained from the implicit dispersion equation
\begin{align}
\det(M(E )-e^{ik}\openone)=0.
\label{eq:implicitdisp}
\end{align}
In systems with some internal periodicity, one can furthermore either restrict the
segment to a single unit cell, or interpret the result as a folded band structure in the reduced Brillouin zone. The result is then identical to the eigenvalues of the corresponding Bloch Hamiltonian
\begin{align}
H(k)=H+we^{-ik}\Delta^{(1N)} +we^{ik}\Delta^{(N1)},
\label{eq:hbloch}
\end{align}
where $\Delta^{(kl)}_{nm}=\delta_{kn}\delta_{lm}$.

As Eq.~\eqref{eq:realtoprop} is a unitary transformation, the eigenvalues of $\mathcal{M}$ and $M$ are identical, so that the condition \eqref{eq:periodiccond} also applies to the propagating-wave basis, whilst the  implicit dispersion equation \eqref{eq:implicitdisp} takes the analogous form
\begin{align}
\det(\mathcal{M}(E )-e^{ik}\openone)=0.
\end{align}
From the scattering perspective, periodic boundary conditions are more intricate. The functions $\lambda_{l}(E)$ appearing in Eq.~\eqref{eq:periodiccond} are then obtained from the condition
\begin{equation}
\det[ S(E )-\sigma_x\,\mathrm{diag}\,(\lambda,1/\lambda)]=0,
\end{equation}
whilst the implicit dispersion equation \eqref{eq:implicitdisp}
can then be written as
\begin{equation}
\det[ S(E )-\sigma_x\,\mathrm{diag}\,(e^{ik},e^{-ik})]=0.
\label{eq:speriodic}
\end{equation}
To verify this relation, we note that starting from Eq. \eqref{eq:mw1},
we can express
\begin{align}
&
S-\sigma_x\,\mathrm{diag}\,(e^{ik},e^{-ik})
\nonumber\\
&
=
-\begin{pmatrix}   i & e^{-ik} \\ e^{ik} & i \end{pmatrix}
\frac{1}{\openone +w  \begin{pmatrix}   -i & e^{-ik} \\ e^{ik} & -i \end{pmatrix}
	(E-H(k))^{-1}_{\{1,N\}}
}
,
\end{align}
where we
used the exact Dyson equation
\begin{align}
[(E-H)^{-1}_{\{1,N\}}]^{-1}=&[(E-H(k))^{-1}_{\{1,N\}}]^{-1}
\nonumber\\
&
+w\,\mathrm{diag}\,(e^{ik},e^{-ik})
.
\end{align}
It follows that the solutions
Equation \eqref{eq:speriodic} indeed coincide with the eigenvalues of the Bloch Hamiltonian \eqref{eq:hbloch}.

\subsection{Quasi-bound states}
Quasi-bound states are defined as solutions without incoming wave components. This is most straightforwardly  formulated for the propagating-wave transfer matrix, which then has to fulfill
\begin{align}
\mathcal{M}_{22}(E )=0.
\label{eq:qbound}
\end{align}
As for the bound states in a closed system with open boundary conditions, the solutions $E _n$ are generally discrete and complex. In the special case of a real-valued solution, $\mathrm{Im}\, E _n=0$, the solution can be interpreted as a stationary lasing state.

In terms of the real-space transfer matrix, the rule \eqref{eq:realtopropelements} gives the condition
\begin{align}
M_{11}(E )+M_{22}(E )+iM_{12}(E )-iM_{21}(E )=0.\label{implicit:quasi}
\end{align}
Furthermore, expression  \eqref{eq:smat} implies that the quasibound-state energies coincide with the poles of the scattering matrix, which can be conveniently expressed as
\begin{equation}
\det S^{-1}(E ) =0.
\end{equation}
Equation \eqref{eq:mw2maintext} furthermore shows that these poles coincide with the eigenvalues of the effective Hamiltonian $H_\mathrm{eff}$ given in Eq. \eqref{eq:heffmain}.

\subsection{Coherent perfect absorption}
For coherent perfect absorption (CPA), we require a stationary state with purely incoming boundary conditions. This is the time-reversed of a stationary lasing state, which fulfills the equivalent conditions
\begin{align}
&\mathcal{M}_{11}(E )=0, \\
& M_{11}(E )+M_{22}(E )+iM_{21}(E )-iM_{12}(E )=0,\label{implicit:cpa}
\\
& \det S(E )=0,
\end{align}
at an energy $E $ that has to be real.
This implies in particular that coherent perfect absorption is related to the \emph{zeros} of the scattering matrix, which are defined as the energies where at least one of its eigenvalues
vanishes.

\subsection{Reflectionlessness, transparency, and invisibility}
By definition of the transport coefficients, the system is reflectionless from the left or right if
\begin{align}
r(E )=0\quad\mbox{ or }\quad r'(E )=0.
\end{align}
For the transfer matrix, this can be written as
\begin{align}
&\mathcal{M}_{21}(E )=0\quad\mbox{ or }\quad \mathcal{M}_{12}(E )=0.
\label{implicit:Invisibility I}
\end{align}

Furthermore, the system is transparent when a probing wave passes through with the same  phase shift $i^N$ as if it was replaced by a lead segment of the same length. Therefore, depending on the side from which the system is probed,
we have
\begin{align}
t(E )=i^N\quad\mbox{ or }\quad t'(E )=i^N,
\end{align}
which for the transfer matrices amounts to
\begin{align}
&
\frac{\mathcal{M}_{22}(E )}{\text{Det}(\mathcal{M})}
=(-i)^N\quad\mbox{ or }\quad \mathcal{M}_{22}(E )=(-i)^N.\label{implicit:invisibility II}
\end{align}
We note that $\frac{\mathcal{M}_{22}(E )}{\text{Det}(\mathcal{M})}=(\mathcal{M}^{-1}(E ))_{11}$, which confirms that the conditions respect symmetry when one reformulates the transfer matrix by iteration from the right to the left lead.
In the Table, for conciseness we specify the condition in this alternative form.

By taking the modulus of these conditions on both sides, they can be relaxed to transparency in terms of the intensity, only.

Finally, we note that a system is \emph{invisible to a source} placed on a given side when it is both reflectionless and transparent under illumination from that side, while it is \emph{invisible to a detector} placed on a given side when it is reflectionless under illumination from that side and transparent under illumination from the other side.


%

\end{document}